\documentclass[a4paper,12pt]{article}

\usepackage{amsmath}
\usepackage[psamsfonts]{amssymb}
\usepackage{rsfs}
\usepackage{bm}

\usepackage{cite}
\usepackage[dvips]{graphicx}
\usepackage{color}

\makeatletter
\@addtoreset{equation}{section}
\makeatother

\begin{document} 

\begin{titlepage}

\baselineskip 10pt
\hrule 
\vskip 5pt
\leftline{}
\leftline{Chiba Univ. Preprint
          \hfill   \small \hbox{\bf CHIBA-EP-171}}
\leftline{\hfill   \small \hbox{May 2008}}
\vskip 5pt
\baselineskip 14pt
\hrule 
\vskip 1.0cm
\centerline{\Large\bf 
} 
\vskip 0.3cm
\centerline{\Large\bf  
Magnetic monopoles and center vortices 
}
\vskip 0.3cm
\centerline{\Large\bf  
as gauge-invariant topological defects  
}
\vskip 0.3cm
\centerline{\Large\bf  
simultaneously responsible for confinement
}

\vskip 0.5cm

\centerline{{\bf 
Kei-Ichi Kondo,$^{\dagger,{1}}$  
}}  
\vskip 0.5cm
\centerline{\it
${}^{\dagger}$Department of Physics, Graduate School of Science, 
}
\centerline{\it
Chiba University, Chiba 263-8522, Japan
}
\vskip 1cm

\begin{abstract}
We give a gauge-invariant definition of the vortex surface in $SU(N)$ Yang-Mills theory without using the gauge fixing procedure.  In this construction, gauge-invariant magnetic monopoles with  fractional magnetic charges emerge in the boundary of the non-oriented vortex surface  such that the asymptotic string tension reproduces the correct $N$-ality dependence.  
We show that gauge-invariant magnetic monopoles and vortices are simultaneously responsible for quark confinement in four dimensional spacetime based on the Wilson criterion. 
These results are extracted from a non-Abelian Stokes theorem derived in the previous paper. 
\end{abstract}

Key words:   magnetic monopole, vortex, Wilson loop, non-Abelian Stokes theorem, quark confinement, N-ality, Yang-Mills theory,  
 
\vskip 0.5cm

PACS: 12.38.Aw, 12.38.Lg 
\hrule  
\vskip 0.1cm
${}^1$ 
  E-mail:  {\tt kondok@faculty.chiba-u.jp}

\par 
\par\noindent


\vskip 0.5cm

\newpage
\pagenumbering{roman}




\end{titlepage}


\pagenumbering{arabic}

\baselineskip 14pt


\section{Introduction}

In understanding the non-perturbative phenomena in the infrared sector of Yang-Mills theory \cite{YM54} and QCD such as quark confinement \cite{Wilson74}, chiral symmetry breaking and $U(1)_A$ problem, some of topological configurations  are believed to play the key role as the dominant dynamical degrees of freedom. 
Examples are magnetic monopoles, center vortices, instantons, merons, etc.
Among them, chiral symmetry breaking and the $U_A(1)$ problem can be explained e.g., by the Yang-Mills instantons \cite{tHooft76,CDG78}, although magnetic monopoles are not excluded as their mechanisms, see e.g. \cite{KK05}. 

Quark confinement is believed to be explained by the condensation of  Abelian magnetic monopoles \cite{dualsuper} and/or  center vortices \cite{center-vortices,Tomboulis93}, since they realize the dual superconductor picture  of QCD vacuum as the most promising scenario for confinement. 
Magnetic monopoles are topological objects of codimension 3: points for $D=3$ and closed loops for $D=4$ in $D$-dimensional spacetime. On the other hand, center vortices are topological objects of codimension 2:  closed loops for $D=3$ and closed surfaces for $D=4$ \cite{center,BFGO99,ER00}.  
For example, the vortex in $D=3$ is in the first step identified with a closed thin tube of magnetic flux which can be thought of as the magnetic field generated by a toroidal solenoid in the limit of vanishing cross section, although the vortex must have a finite transverse extension to correctly reproduce the adjoint string tension and yield the finite action of vortices \cite{Cornwall98}. 
Recent lattice simulations exhibit both Abelian magnetic monopole dominance \cite{SNW94} and center vortex dominance \cite{center} for the string tension. 
Moreover, if either Abelian magnetic monopoles or center vortices in Yang-Mills theory are removed from the ensemble of configurations, confinement is found to be lost.  Moreover, chiral symmetry breaking is also lost. 
It is known according to numerical simulations that magnetic monopoles and vortices are strongly correlated. 
Incidentally, merons \cite{AFF76} can be also a candidate for confiners \cite{CDG78} and have something to do with magnetic monopoles due to numerical simulations on a lattice \cite{meron}. 
For recent reviews, see e.g., \cite{CP97} for Abelian magnetic monopole and \cite{Greensite03} for center vortices.

Yang-Mills instantons and merons are respectively self-dual Euclidean and non-self-dual Euclidean/Minkowski solutions of the gauge covariant Yang-Mills  field equations derived from the Yang-Mills action.  
However, Abelian magnetic monopoles \cite{tHooft81} or center vortices \cite{center-vortices} in $SU(N)$ Yang-Mills theory have been obtained as gauge fixing defects by a partial gauge fixing  $G=SU(N) \rightarrow H$  where the gauge degrees of freedom is used to transform the gauge field variable (link variable on a lattice) as close as possible to a subgroup $H$ which is left unbroken:  the maximal torus (Cartan) subgroup $H= U(1)^{N-1}$ for the maximal Abelian gauge \cite{tHooft81,KLSW87} or the center subgroup $H=Z(N)$ for the maximal center gauge.  
Therefore, the current method of constructing Abelian magnetic monopoles and center vortices could not escape the charge of the gauge artifact.

In a series of recent papers, we have succeeded to give a gauge-invariant description of the dual superconductivity in Yang-Mills theory in the continuum \cite{KMS06,KSM08} and on a lattice \cite{KKMSSI05,IKKMSS06,SKKMSI07,SKKMSI07b,KSSMKI08} by developing the approach founded in \cite{Cho80,FN98,Shabanov99}.  
Especially, the Wilson loop operator is expressed exactly in terms of a gauge-invariant magnetic current \cite{Kondo08} and the magnetic monopole can be defined in a gauge-invariant way according to a non-Abelian Stokes theorem for the Wilson loop operator \cite{DP89,DP96,KondoIV,Kondo08}. 
These results enable us to clarify the role of magnetic monopole in confinement. 

In this paper, we give a gauge-invariant definition of a vortex  which can play the same role as the center vortex to sweep away distrust of gauge artifact. 
Indeed, this is achieved without relying on the (partial) gauge fixing such as the maximal Abelian gauge and the maximal center gauge. 
Consequently, the gauge-invariant magnetic monopoles emerge in the boundary of the closed vortex surface in $D=4$. 
We show that both magnetic monopoles and vortices are necessary from a viewpoint of quark confinement based on the Wilson criterion such that the asymptotic string tension reproduces the correct $N$-ality dependence. 
The result of this paper shows that the vortex and the magnetic monopole can be the alternative view of the one and the same fundamental dynamics in Yang-Mills theory defined in a gauge-invariant manner.

\section{Wilson loop operator and magnetic monopoles}\label{sec:NAST}
 
For the Yang-Mills connection (one-form) $\mathscr{A}=\mathscr{A}_\mu(x) dx_\mu=\mathscr{A}^A_\mu(x)T_A dx_\mu$ for a gauge group $G$, the Wilson loop operator $W_C[\mathscr{A}]$ along a closed loop $C$ is defined by
\begin{equation}
 W_C[\mathscr{A}] :=  {\rm tr} \left[ \mathscr{P} \exp \left\{ ig \oint_C \mathscr{A} \right\} \right]/{\rm tr}(\mathbf{1}) ,
\end{equation}
where $\mathscr{P}$ denotes the path-ordering prescription. 
The non-Abelian Stokes theorem (NAST) enables us to rewrite the Wilson loop operator into the surface integral form over the surface $\Sigma$ bounding $C$ ($\partial \Sigma=C$).
A version of the non-Abelian Stokes theorem without any  path or surface ordering is known as the Diakonov-Petrov version \cite{DP89} of a non-Abelian Stokes theorem. 
The Diakonov-Petrov version of NAST was originally derived in  \cite{DP89} for $G=SU(2)$ case and later developed and extended to $G=SU(N)$ case in \cite{DP96,KondoIV}. 
Moreover, it has been shown \cite{Kondo08}  that the Wilson loop operator is rewritten in terms of two gauge-invariant conserved currents, the  ``magnetic-monopole current''   $k$ and the ``electric current'' $j$, defined by applying the exterior derivative $d$, the coderivative (adjoint derivative) $\delta$ and Hodge star operation $*$ to $f$: 
\begin{align}
 k:=  \delta *f = *df, \quad
 j:= \delta f 
  ,
  \label{def-k-j}
\end{align}
where  $f$ is the gauge-invariant two-form defined from the $SU(N)$ gauge connection $\mathscr{A}_\mu(x)=\mathscr{A}^A_\mu(x)T_A$ by 
\begin{align}
 f_{\mu\nu}(x) 
  =&  \partial_\mu 2{\rm tr}(\bm{n}(x) \mathscr{A}_\nu(x)) - 
  \partial_\nu 2{\rm tr}(\bm{n}(x) \mathscr{A}_\mu(x)) 
  \nonumber\\
  &+ 2{\rm tr}(\frac{2(N-1)}{N}ig^{-1} \bm{n}(x) [\partial_\mu \bm{n}(x), \partial_\nu \bm{n}(x)])  
   ,
\end{align}
using the $su(N)$ Lie-algebra valued field $\bm{n}(x)= n_A(x) T_A$ called the (normalized) color field  defined by
\begin{align}
 \bm{n}(x) = n_A(x) T_A := \sqrt{\frac{2N}{N-1}}\xi(x) \mathcal{H} \xi^\dagger(x) ,\quad \xi(x) \in G/\tilde{H} 
   ,
\end{align}
with the stability group $\tilde H$ specified later  
and $\mathcal{H}$ given by 
\begin{equation}
 \mathcal{H} := 
\bm{\Lambda} \cdot \bm{H}  = \sum_{j=1}^{r} \Lambda_j H_j 
 , 
\end{equation}
where $H_j$ ($j=1,, \cdots, r$) are the generators from the Cartan subalgebra of $su(N)$ ($r=N-1$ is the rank of the gauge group $G=SU(N)$) and 
$r$-dimensional vector $\Lambda_j$ ($j=1,, \cdots, r$) is the highest weight of the representation in which the Wilson loop is considered. 
Indeed, both currents are conserved in the sense that 
$\delta k=0$ and $\delta j=0$.

For the Wilson loop operator in the fundamental representation of $G=SU(N)$, $\tilde{H}=U(N-1)$ \cite{Kondo08} and the Wilson loop operator is rewritten as 
\begin{align}
 W_C[\mathscr{A}] 
 =& \int [d\mu(\xi)]_{\Sigma} \exp \left[ i \sqrt{\frac{N-1}{2N}} g \int_{\Sigma} f \right]
 \nonumber\\
=& \int [d\mu(\xi)]_{\Sigma} \exp \left\{ i  \sqrt{\frac{N-1}{2N}} g(k, \Xi_{\Sigma}) + i  \sqrt{\frac{N-1}{2N}}  g(j, N_{\Sigma}) \right\} ,
\label{NAST-SUN}
\end{align}
where $[d\mu(\xi)]_{\Sigma}$ is the product of the Haar measure $d\mu(\xi)$ on $SU(N)/U(N-1)$ over $\Sigma$:
\begin{align} 
 [d\mu(\xi)]_{\Sigma}  := \prod_{x \in \Sigma} d\mu(\xi(x))
  ,
\end{align}
and $\Xi_{\Sigma}$ is the $(D-3)$-form and $N_{\Sigma}$ is the one-form defined by
\begin{equation}
 \Xi_{\Sigma} := * d\Theta_{\Sigma} \Delta^{-1} = \delta *\Theta_{\Sigma} \Delta^{-1} , \quad
 N_{\Sigma} := \delta \Theta_{\Sigma} \Delta^{-1}
 ,
\end{equation}
with the $D$-dimensional Laplacian (or the d'Alembertian in the Minkowski spacetime) $\Delta:=d\delta+\delta d$ and   the two-form $\Theta_{\Sigma}$  called the vorticity tensor as an antisymmetric tensor of rank two:
\begin{align}
 \Theta^{\mu\nu}_{\Sigma}(x) 
:=   \int_{\Sigma}  dS^{\mu\nu}(X(\sigma)) \delta^D(x-X(\sigma)) 
 .
\end{align}
Note that $k$ and $\Xi_{\Sigma}$ are $(D-3)$-forms, while  $j$ and $N_{\Sigma}$ are one-forms for any $D$ in $D$ dimensional spacetime:  
\begin{align}
  (k, \Xi_{\Sigma}) 
:= \frac{1}{(D-3)!} \int d^Dx k^{\mu_1 \cdots \mu_{D-3}}(x) \Xi^{\mu_1 \cdots \mu_{D-3}}_{\Sigma}(x) ,
\quad
  (j, N_{\Sigma}) 
:=  \int d^Dx j^{\mu}(x) N^{\mu}_{\Sigma}(x) 
 .
\end{align}

For $G=SU(2)$, in particular, arbitrary representation is characterized by an integer or a half-integer $J=\frac12, 1, \frac32, 2, \frac52, \cdots$. The Wilson loop operator in the representation $J$ of $SU(2)$ obey the non-Abelian Stokes theorem:
\begin{equation}
 W_C[\mathscr{A}] 
= \int [d\mu(\xi)]_{\Sigma} \exp \left\{ i J g (k, \Xi_{\Sigma}) + iJ g  (j, N_{\Sigma}) \right\} 
.
\end{equation}
This agrees with (\ref{NAST-SUN}) for a fundamental representation $J=\frac12$ of SU(2).

We focus on the magnetic contribution $W_C^m$ defined by 
\begin{equation}
 W_C^m    
= \exp \left\{ i  \sqrt{\frac{N-1}{2N}} g  (k, \Xi_{\Sigma})  \right\} 
 .
\end{equation}
For $D=3$, it has been shown \cite{Kondo08} that the magnetic charge $q_m$ defined by $q_m=\int d^3x k^0$ for $SU(2)$ obeys the  quantization condition:    
\begin{equation}
  q_m=4\pi g^{-1} n, \quad n \in \mathbb{Z} = \{ \cdots, -2,  - 1, 0, +1, +2, \cdots \} 
   .
   \label{qc0}
\end{equation}
This follows from the condition that the non-Abelian Stokes theorem should not depend on the surface $\Sigma$ chosen for  bounding the loop $C$, since the original Wilson loop is defined for the specified closed loop $C$.
For an ensemble of point-like magnetic charges located at $x=z_a$ ($a=1, \cdots, n$)
\begin{equation}
 k^0(x) =   \sum_{a=1}^{n} q_m^a \delta^{(3)}(x-z_a) 
  ,
  \quad 
  q_m^a=4\pi g^{-1} n_a , \quad n_a \in \mathbb{Z} 
   ,
\end{equation}
we have a geometric representation:
\begin{equation}
W_C^m  = \exp \left\{ i \frac12 \frac{g}{4\pi} \sum_{a=1}^{n} q_m^a \Omega_{\Sigma}(z_a)  \right\}
= \exp \left\{ i \frac12  \sum_{a=1}^{n} n_a \Omega_{\Sigma}(z_a)  \right\}
 , \quad n_a \in \mathbb{Z} 
  ,
\end{equation}
where $\Omega_{\Sigma}(x)$ is the the solid angle  under which the surface $\Sigma$ shows up to an observer at the point $x$. 
Therefore, a magnetic monopole with a unit magnetic charge $q_m=4\pi g^{-1}$ in the neighborhood of the surface $\Sigma$ gives a non-trivial factor $\exp [ \pm i \pi]=-1$ to the Wilson loop operator $W_C^m$, since 
$\Omega_{\Sigma}(z_a)=\pm 2\pi$ when $z_a$ is just below and above  the surface $\Sigma$. 
This is a nice feature of magnetic monopole for explaining quark confinement based on the Wilson loop. 

\begin{figure}[ptb]
\begin{center}
\includegraphics[height=4.0cm,clip]{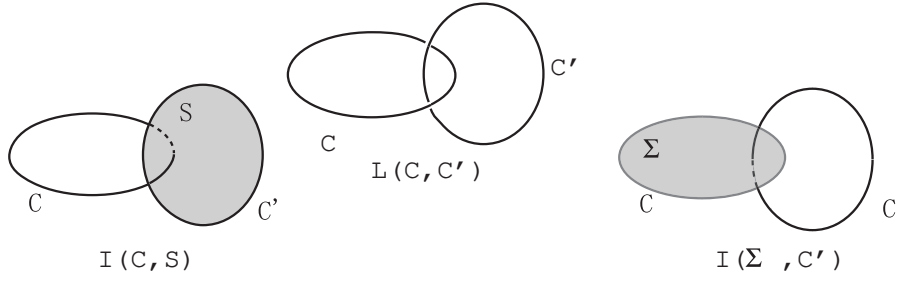}
\end{center} 
 \caption[]{
  In $D=3$ dimensional space-time,  a non-trivial contribution to the Wilson loop operator comes from    
 (Center panel) 
 the linking $L(C,C^\prime)$ between the closed loop $C$ and a closed (magnetic current) loop $C^\prime$ as a boundary of the vortex sheet $S$.
(Left panel) intersection $I(C=\partial \Sigma,S)$  between the closed loop $C$ and an open vortex sheet $S$ bounding the closed loop $C^\prime$, 
or 
(Right panel) intersection $I(\Sigma,C^\prime=\partial S)$ between the open surface $\Sigma$ bounding the closed loop $C$ and a closed loop $C^\prime$. Three are equivalent descriptions: 
$L(C,C^\prime)=I(C=\partial \Sigma,S)=I(\Sigma,C^\prime=\partial S)$.
}
 \label{fig:vortex3d}
\end{figure}

For $D=4$, however, we fall in a trouble about the magnetic monopole just defined. 
For $D=4$, an ensemble of magnetic currents on closed loops $C^\prime_a$ ($a=1,\cdots,n$):
\begin{equation}
 k^\mu(x) = \sum_{a=1}^{n} q_m^a \oint_{C^\prime_a} dy^\mu_a \delta^{(4)}(x-y_a) , 
 \quad 
 q_m^a = 4\pi g^{-1} n_a 
  ,
\end{equation}
leads to \cite{Kondo08}
\begin{equation}
 W_C^m    
= \exp \left\{ i \frac12 g \sum_{a=1}^{n} q_m^a L(C^\prime_a, \Sigma)  \right\}
= \exp \left\{  2\pi i \sum_{a=1}^{n} n_a L(C^\prime_a, \Sigma)  \right\}
 , 
\quad n_a \in \mathbb{Z} 
  ,
\end{equation}
where
$L(C^\prime,\Sigma)$ is the linking number between the curve $C^\prime$ and the surface $\Sigma$:
\begin{equation}
  L(C^\prime, \Sigma)  = L(\Sigma,C^\prime) 
  := \oint_{C^\prime} dy^\mu(\tau) \Xi^\mu_{\Sigma}(y(\tau)) 
  ,
\end{equation} 
where the curve $C^\prime$ is identified with the trajectory of a magnetic monopole and the surface $\Sigma$ with the world sheet of a hadron (meson) string representing a quark-antiquark pair.  For $D=3$ case, see Fig.~\ref{fig:vortex3d}.
However, such magnetic loops carrying the magnetic charge obeying the quantization condition (\ref{qc0}) do not give non-trivial contributions to the Wilson loop, since $n_a$ and $L$ are integers. 
If the quantization condition (\ref{qc0}) is true, the magnetic monopole can not be the topological defects responsible for quark confinement. 
In the following, we discuss how this dilemma is resolved.

\section{Magnetic monopole and vortex}\label{sec:a}

In the following, we extensively use the techniques developed by Engelhardt and Reinhardt \cite{ER00} in constructing a continuum analogue of the maximal center gauge and center projection, but from a different angle in this paper.

We first consider the $D=4$ case. 
Suppose that the magnetic current $k^\mu$ has the support on the closed loop $C^\prime$ in $D=4$ dimensions:
\begin{align}
 k^\mu(x) = k^\mu(x;C^\prime) := \Phi \oint_{C^\prime} dy^\mu \delta^4(x-y) 
  ,
\end{align}
where  
$\Phi$ is a real number representing the magnetic flux carried by the magnetic charge $q_m$ to be discussed later in detail. 
The magnetic charge $q_m$ is defined by 
\begin{align}
 q_m=\int d^3 \tilde{\sigma}_\mu k^\mu 
  ,
\end{align}
where $\bar{x}^\mu$ denotes a parameterization of the 3-dimensional volume $V$ and 
 $d^3 \tilde{\sigma}_{\mu} $ is the dual of the 3-dimensional volume element $d^3 \sigma^{\gamma_1\gamma_2\gamma_3}$:
\begin{align}
 d^3 \tilde{\sigma}_{\mu} 
:=  \frac{1}{3!}   \epsilon_{\mu\gamma_1\gamma_2\gamma_3} d^3 \sigma^{\gamma_1\gamma_2\gamma_3}
 , \quad
 d^3 \sigma^{\gamma_1\gamma_2\gamma_3}
 :=  \epsilon_{\beta_1\beta_2\beta_3} 
 \frac{\partial \bar{x}^{\gamma_1}}{\partial \sigma_{\beta_1}} \frac{\partial \bar{x}^{\gamma_1}}{\partial \sigma_{\beta_1}}  \frac{\partial \bar{x}^{\gamma_1}}{\partial \sigma_{\beta_1}}  d\sigma_{1} d\sigma_{2} d\sigma_{3} 
  .
\end{align}

First of all, we look for the field strength ${}^*f^{\mu\nu}(x;S)$ with the support $S$, a two-dimensional surface bounding the closed loop $C^\prime$, $\partial S=C^\prime$, so that  (\ref{def-k-j}) holds:
\begin{align}
  \partial_\nu {}^*f^{\mu\nu}(x;S)  = k^\mu(x;C^\prime) 
  .
\end{align}
Such a solution is given by
\begin{align}
 {}^*f^{\mu\nu}(x;S) 
 = \Phi \int_{S:\partial S=C^\prime} d^2 \sigma^{\mu\nu} \delta^4(x-\bar{x}(\sigma)) 
  .
\end{align}
In fact, it satisfies the equation:
\begin{align}
 \partial_\nu  {}^*f^{\mu\nu}(x;S) 
 =& \Phi \int_{S} d^2 \sigma^{\mu\nu} \partial_\nu^{x}   
\delta^4(x-\bar{x}(\sigma)) 
\nonumber\\
 =&  \Phi \int_{S} d^2 \sigma^{\mu\nu} \partial_\nu^{\bar{x}}   \delta^4(x-\bar{x}(\sigma)) 
\nonumber\\
 =&  \Phi \oint_{\partial S=C^\prime} d \bar{x}^{\mu}   \delta^4(x-\bar{x}(\sigma)) 
  ,
\end{align}
where we have used the Stokes theorem in the last step.

Next, we proceed to obtain the gauge potential $b_\mu(x;V)$ giving the field strength $f^{\mu\nu}(x;S)$ just obtained, i.e.,  
\begin{align}
 f_{\mu\nu}(x;S) =& \partial_\mu b_\nu(x;V) - \partial_\nu b_\mu(x;V) 
 , 
\end{align}
such that $b_\mu(x;V)$ has the support only on the open set $V$, the three-dimensional volume
\footnote{
The precise position of the open set $V$ is irrelevant for the value of the Wilson loop as shown below. 
In fact, the open set $V$ can be deformed arbitrarily by singular gauge transformations in such a way that its boundary $\partial V$ representing the position of the magnetic flux of the vortex is fixed. 
}
whose boundary is $S$: $\partial V=S$.
Note that ${}^*f^{\mu\nu}(x;S) $ is cast into  
\begin{align}
 {}^*f^{\mu\nu}(x;S) 
 =& \Phi \int_{V:\partial V=S} d^3 \sigma^{\mu\nu\kappa}  \partial_\kappa^{\bar{x}}  \delta^4(x-\bar{x}(\sigma)) 
\nonumber\\
 =& \Phi \epsilon^{\mu\nu\alpha\beta} \frac{1}{3!} \epsilon_{\beta\gamma_1\gamma_2\gamma_3} \int_{V:\partial V=S} d^3 \sigma^{\gamma_1\gamma_2\gamma_3}  \partial_\alpha^{x}  \delta^4(x-\bar{x}(\sigma)) 
\nonumber\\
 =&  \epsilon^{\mu\nu\alpha\beta} \partial_\alpha^{x}   [\Phi  \frac{1}{3!} \epsilon_{\beta\gamma_1\gamma_2\gamma_3} \int_{V:\partial V=S} d^3 \sigma^{\gamma_1\gamma_2\gamma_3}  \delta^4(x-\bar{x}(\sigma)) ]
\nonumber\\
 =& \frac12 \epsilon^{\mu\nu\alpha\beta} \{ \partial_\alpha^{x}   [\Phi   \int_{V:\partial V=S} d^3 \tilde{\sigma}_{\beta}  \delta^4(x-\bar{x}(\sigma))] - (\alpha \leftrightarrow \beta) \}
  ,
\end{align}
where we have used the Gauss (Stokes) theorem in the first equality.
 Therefore, the gauge potential $b_\mu(x;V)$ is determined up to a gauge transformation:
\begin{align}
   b_\mu(x;V) =  \Phi   \int_{V:\partial V=S} d^3 \tilde{\sigma}_{\mu}  \delta^4(x-\bar{x}(\sigma))
    .
\end{align}
This corresponds to an explicit (singular) gauge field representation of an {\it ideal vortex configuration} in 4 space-time dimensions \cite{ER00}.

In $D$ dimensions, the magnetic current $k$ is $(D-3)$-form with the support on the $(D-3)$-dimensional subspace $C^\prime_{D-3}$ in $D$-dimensional spacetime:
\begin{align}
 k(x) = k(x;C^\prime_{D-3}) := \Phi \oint_{C^\prime_{D-3}} d^{D-3}\bar{x}(\sigma) \delta^D(x-\bar{x}(\sigma)) 
  .
\end{align}
Then it is not difficult to show that an ideal vortex configuration should read
\begin{align}
 b_\mu(x;V) =& \Phi   \int_{V_{D-1}:\partial V_{D-1}=S_{D-2}} d^{D-1} \tilde{\sigma}_{\mu} \ \delta^D(x-\bar{x}(\sigma))
  .
\end{align}
The gauge field $b_\mu(x;V)$ for the vortex is not unique as mentioned above. 
Actually, the ideal vortex field $b_\mu(x;V)$ can be gauge transformed to a {\it thin vortex field} $a_\mu(x;S)$ which has the support only on the boundary $S_{D-2}=\partial V_{D-1}$ of $V_{D-1}$:  
\begin{equation}
  b_\mu(x;V)= a_\mu(x;S) + iU(x;V) \partial_\mu U^\dagger(x;V) 
  ,
  \label{gt}
\end{equation}
so that 
$a_\mu(x;S)$ carries the same flux located on $S$ as that carried by $b_\mu(x;V)$.
We adopt the gauge transformation: 
\begin{equation}
  U(x;V) = \exp \left[ i\Phi \Omega_V(x) \right] 
   ,
\end{equation}
where $\Omega_V(x)$ is the solid angle taken up by the volume $V$ viewed from $x$:
\begin{equation}
   \Omega_V(x) := \frac{1}{A_{D-1}} \int_{V_{D-1}} d^{D-1} \tilde{\sigma}_\mu \ \frac{\bar{x}(\sigma)-x_\mu}{(\bar{x}(\sigma)-x_\mu)^D} ,
   \quad 
  A_{D-1} := \frac{2\pi^{D/2}}{\Gamma(D/2)} 
   ,
\end{equation}
with $A_{D-1}$ being the area of unit sphere $S^{D-1}$ in $D$ dimensions.  
The solid angle $\Omega_V(x)$ defined in this way is normalized to unity for a point $x$ inside the volume $V$.  
Note that the solid angle is defined with a sign depending on the orientation of $V$ as rays emanating from $x$ pierce $V$. 
A deformation of $V$ keeping its boundary $S$ fixed leaves the solid angle invariant unless $x$ crosses $V$.  When $x$ crosses $V$, the solid angle changes by an integer. 
\footnote{
Whether the flux of a vortex is electric or magnetic depends on the position of the $D-2$ dimensional vortex surface $S$ in $D$-dimensional spacetime. For example, in $D=4$ the vortex defined by the boundary of a purely spatial 3-dimensional volume $V$ carries only electric flux, which is directed normal to the vortex surface $S=\partial V$.  On the other hand, a vortex $S=\partial V$ defined by a volume $V$ evolving in time represents at a fixed time a closed loop and carries the magnetic flux, which is tangential to the vortex loop \cite{ER00}. 
}

Now we show that the thin vortex field is obtained in the form:
\begin{align}
 a_\mu(x;S) = \Phi   \int_{S_{D-2}=\partial V_{D-1}} d^{D-2} \tilde{\sigma}_{\mu\lambda} \ \partial_\lambda D(x-\bar{x}(\sigma))
  ,
  \label{tv-field}
\end{align}
where  $D(x-\bar{x}(\sigma))$ is the Green function of the $D$-dimensional Laplacian defined by
\begin{align}
  - \partial_\mu \partial_\mu D(x-\bar{x}(\sigma)) = \delta^D(x-\bar{x}(\sigma))
  ,
\end{align}
with the explicit form:
\begin{align}
 D(x-y)
 = \frac{\Gamma(D/2-1)}{4\pi^{D/2}[(x-y)^2]^{(D-2)/2}} 
  .
\end{align}
The ideal vortex field $b_\mu(x;V)$ has support only on $V$ and hence it vanishes outside the vortex $V$, i.e., $b_\mu(x;V)=0$ for $x \notin V$.  Therefore,  outside the vortex $V$, i.e., $x \in \mathbb{R}^D - V$, the thin vortex field $a_\mu(x;S)$ is the pure gauge due to (\ref{gt}):
\begin{equation}
  a_\mu(x;S) =  - iU(x;V) \partial_\mu U^\dagger(x;V) , \quad x \notin V
  ,
\end{equation}
which implies the vanishing field strength $f_{\mu\nu}(x)=0$ outside the vortex $V$.  This is reasonable, because the magnetic flux is contained in the vortex sheet $S$.
The magnetic field computed from the curl of $a_\mu(x;S)$ can be localized on $C^\prime=\partial S$.

The derivation of (\ref{gt}) and (\ref{tv-field}) is as follows.
By using the fact that the solid angle is rewritten as
\begin{equation}
   \Omega_V(x) := \int_{V_{D-1}} d^{D-1} \tilde{\sigma}_\mu \partial_\mu^{x} D(x-\bar{x}(\sigma)) 
   ,
\end{equation}
we obtain
\begin{align}
 iU(x;V) \partial_\mu U^\dagger(x;V) 
  =   \Phi \partial_\mu \Omega_V(x)
  =  \Phi \int_{V_{D-1}} d^{D-1} \tilde{\sigma}_\nu \partial_\mu \partial_\nu  D(x-\bar{x}(\sigma)) 
 .
\end{align}
We have the decomposition:
\begin{align}
  \partial_\mu \partial_\nu
  =  \delta_{\mu\nu} \partial^2 - ( \delta_{\mu\nu} \partial^2 - \partial_\mu \partial_\nu)
  = \delta_{\mu\nu} \partial^2 - \frac12 \epsilon_{\mu\rho\alpha\beta} \epsilon_{\nu\sigma\alpha\beta} \partial_\rho \partial_\sigma 
 ,
\end{align}
which is used to rewrite the pure gauge form into 
\begin{align}
& iU(x;V) \partial_\mu U^\dagger(x;V) 
\nonumber\\
  =& \Phi \int_{V_{D-1}} d^{D-1} \tilde{\sigma}_\mu \partial^2  D(x-\bar{x}(\sigma)) 
  - \Phi \int_{V_{D-1}} d^{D-1} \tilde{\sigma}_\nu \frac12 \epsilon_{\mu\rho\alpha\beta} \epsilon_{\nu\sigma\alpha\beta} \partial_\rho \partial_\sigma   D(x-\bar{x}(\sigma))  
\nonumber\\
  =& -\Phi \int_{V_{D-1}} d^{D-1} \tilde{\sigma}_\mu  \delta^D(x-\bar{x}(\sigma)) 
  - \Phi \int_{\partial V_{D-1}} d^{D-2} \tilde{\sigma}_{\mu\lambda}  \partial_\lambda   D(x-\bar{x}(\sigma))  
 ,
\end{align}
where we have used the definition of the Green function and the Stokes theorem in the last equality. 
In other words, the thin vortex field $a_\mu(x;S)$ is the transverse part of the ideal vortex $b_\mu(x;V)$.

\begin{figure}[ptb]
\begin{center}
\includegraphics[height=4cm,clip]{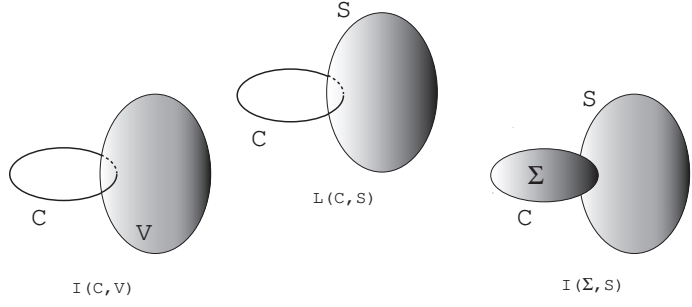}
\end{center} 
 \caption[]{
 In $D=4$ dimensional case,  
(Center panel) Linking $L(C,S)$ between the closed loop $C$ and a closed 2-dim. vortex surface $S$ bounding a 3-dim. vortex $V$. 
(Left panel) Intersection $I(C=\partial \Sigma,V)$ between the closed loop $C$ and  the vortex $V$.
(Right panel) Intersection $I(\Sigma,S=\partial V)$ between a minimal surface $\Sigma$ bounded by the closed loop $C$ and a closed  vortex surface $S$ in 4-dimensional space-time. They are equivalent:
$L(C,S)=I(C=\partial \Sigma,V)=I(\Sigma,S=\partial V)$.
}
 \label{fig:linking-d4}
\end{figure}

Finally, we find that the surface integral of $f(x;S)$ over $\Sigma$ bounded by the Wilson loop $C$  is equivalent to the line integral of $a(x;S)$ along the closed loop $C$:
\begin{align}
 \int_{\Sigma} f(x;S) 
=& \int_{\Sigma} db(x;V) 
\nonumber\\
  =&  \oint_{\partial \Sigma=C}  b(x;V) 
\nonumber\\
  =& \oint_{C} [a(x;S) + iU(x;V) d U^\dagger(x;V) ]
\nonumber\\
  =& \oint_{C}  a(x;S) 
 ,
\end{align}
since the contribution from the last term $iU(x;V) d U^\dagger(x;V)=\Phi \partial_\mu \Omega_V(x)$ vanishes for any closed loop $C$.
Then the line integral is cast into 
\begin{align}
  \oint_{C} dx^\mu a_\mu(x;S) 
  =& \int_{\Sigma} d^2 \sigma_{\nu\mu}(x) \partial_\nu a_\mu(x;S) 
\nonumber\\
  =& \Phi \int_{\Sigma} d^2 \sigma_{\nu\mu}(x) \int_{\partial V_{D-1}} d^{D-2} \tilde{\sigma}_{\mu\lambda}  \partial_\nu  \partial_\lambda   D(x-\bar{x}(\sigma)) 
\nonumber\\
  =& \Phi \frac{-1}{4} \int_{\Sigma} d^{D-2} \tilde{\sigma}_{\rho\sigma}(x) \epsilon_{\mu\nu\rho\sigma} \int_{\partial V_{D-1}} d^{2}  \sigma_{\alpha\beta}(\bar{x}) \epsilon_{\mu\lambda\alpha\beta} \partial_\nu  \partial_\lambda   D(x-\bar{x}) 
\nonumber\\
  =& \Phi \frac{-1}{2} \int_{\Sigma} d^{D-2} \tilde{\sigma}_{\alpha\beta}(x)  \int_{\partial V_{D-1}} d^{2}  \sigma_{\alpha\beta}(\bar{x}) \partial^2   D(x-\bar{x}) 
\nonumber\\
  &  + \Phi  \int_{\Sigma} d^{D-2} \tilde{\sigma}_{\beta\sigma}(x)  \int_{\partial V_{D-1}} d^{2}  \sigma_{\alpha\beta}(\bar{x}) \partial_\alpha  \partial_\sigma   D(x-\bar{x}) 
\nonumber\\
  =& \Phi \frac{1}{2} \int_{\Sigma} d^{D-2} \tilde{\sigma}_{\alpha\beta}(x)  \int_{\partial V_{D-1}} d^{2}  \sigma_{\alpha\beta}(\bar{x}) \delta^D(x-\bar{x}) 
,
\end{align}
where we have used the fact that the second term vanishes due to the Stokes theorem $\partial \partial V =0$ for obtaining the last result. 
Therefore, the line integral is rewritten as
\begin{align}
 \oint_{C} dx^\mu a_\mu(x;S) 
 = \Phi  I(\Sigma,S_{D-2}=\partial V_{D-1})  
,
\end{align}
in terms of the intersection number $I(\Sigma,S)$ between the world sheet $\Sigma$ of the hadron string and the vortex sheet $S$.  
It is known that the intersection number $I(\Sigma,S)$ is equal to the linking number $L(C=\partial \Sigma, S=\partial V)$ between the Wilson loop $C$ and the vortex sheet $S$, see Fig.~\ref{fig:linking-d4}:
\begin{align}
I(\Sigma,S_{D-2}=\partial V_{D-1}) =& 
 L(C=\partial \Sigma, S_{D-2}=\partial V_{D-1}) 
\nonumber\\
  =&   \frac12 \int_{\Sigma} d^{D-2} \tilde{\sigma}^{\alpha\beta}(x) \int_{S} d^2 \sigma^{\alpha\beta}(\bar{x}) \delta^D(x-\bar{x})
 .
\end{align}

Thus the thin vortex  contributes to the $SU(2)$ Wilson loop in the fundamental representation:
\begin{align}
\exp \left[ i \frac12 g \int_{\Sigma} f(x;S) \right]
=  \exp \left[ i \frac12 g\Phi L(C, S=\partial V)  \right]
 =  z^{L(C, S=\partial V)}
, \quad
 z := e^{i\frac12 g\Phi} 
 .
\end{align}
In general, $z$ is a complex number of modulus one, i.e., an element of U(1). 
If the magnetic charge obeys the Dirac quantization condition:
\begin{align}
\Phi= 2\pi g^{-1} n \ (n \in \mathbb{Z})
,
\label{qc}
\end{align}
then $z$ reduces to the center element $\mathbb{Z}_2$ of SU(2):
\begin{align}
 z = e^{i\pi n} = \pm 1 , \quad z {\bf 1} \in \mathbb{Z}_2
 .
 \label{center}
\end{align}
The quantization condition (\ref{qc}) will be called the  \textit{fractional} quantization condition which happens to agree with the Dirac one for SU(2), which is realized as a special case of the general quantization condition for $SU(N)$ discussed later. 
If the magnetic charge obeyed the quantization condition
\begin{align}
\Phi= 4\pi g^{-1} n \ (n \in \mathbb{Z})
,
\label{qc2}
\end{align}
then $z$ would be trivial, i.e., $z=1$.
Such a vortex can not give a non-trivial contribution to the Wilson loop.
Therefore, the thin vortex carrying the fractional magnetic charge yields a center element under the fractional magnetic charge quantization (\ref{qc}). 
For $D=4$, $V$ is a three-dimensional volume and its boundary $S$ is a closed two-dimensional surface $S=\partial V$. 
If $S$ is an oriented closed surface, then its boundary $\partial S=C^\prime$ is empty, and hence the magnetic current $k^\mu(x;C^\prime)$ does not exist in this case, since its support $C^\prime=\partial S=\partial \partial V$ vanishes. 
Therefore, for the non-vanishing magnetic current $k_\mu(x;C)$ to exist in the boundary $C^\prime=\partial S$ of $S$, the vortex surface $S=\partial V$ must be non-oriented.  
Is there any relationship between the non-orientedness of the vortex surface and the fractional magnetic charge (\ref{qc})?  We will try to answer this question in the next section.

\section{Non-orientedness of the vortex surface and fractional magnetic charge quantization}\label{sec:topology}

\begin{figure}[ptb]
\begin{center}
\includegraphics[height=4cm,clip]{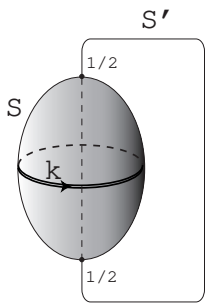}
\end{center} 
 \caption[]{
The schematic view of three-dimensional slice for $D=4$ spacetime dimensions.  A vortex surface $S$ intersects  another surface $S'$ which includes $\Sigma$ bounding $C$.
The vortex surface $S$ is not globally oriented, but consists of patches of different orientation.  Boundaries of such patches are tagged by magnetic monopole (indicated as a thick line around the equator of the surface $S$).
}
 \label{fig:linking-vortex}
\end{figure}

The above considerations suggest that in the continuum theory a smooth vortex surface $S$ consists of surface patches $S_n$ of different orientations  $S=\cup_{n} S_n$, that is to say, the vortex surface $S$ is not globally oriented and that the magnetic monopole loops $k_\mu^n$ emerge at the boundaries $\partial S_n$ of vortex surface patches $S_n$ where the magnetic flux direction on $\partial S_n$ is defined by the orientation of the patch $S_n$ (and vice versa).  
Then we can understand only a fraction of the elementary magnetic charge dictated by the Dirac quantization condition is carried by the magnetic flux $\Phi$ on an isolated patch $S_n$ of the vortex sheet $S$. 
Hence the magnetic current $k_\mu$ of proper magnetic charges satisfying the quantization condition (\ref{qc2}) is reproduced only when the open vortex surface patches $S_n$ are glued together to obtain a non-oriented closed vortex surface $S$ by combining their magnetic loops $k_\mu^n$ together. 
See Fig.~\ref{fig:linking-vortex}.
In fact, if two patches are glued together such that the surface orientation does not change across the boundary (the surface is globally oriented), the magnetic current at the boundary precisely cancel.  
For D=4, thus, the thin vortex defined on a non-oriented closed surface $S$ if linked to the Wilson loop $C=\partial \Sigma$, gives a non-trivial contribution to the Wilson loop average. See Fig.~\ref{fig:linking-d4}

Engelhardt and Reinhardt \cite{ER00} have shown that the continuum center vortex configurations generate the Pontryagin index $\pm 1/2$ as self-intersections of the vortex network.
It is well known in topology that the self-intersection number of closed, {\it globally oriented} two-dimensional surface in $\mathbb{R}^4$ vanishes. 
This implies that the Pontryagin index vanishes for globally oriented vortex surfaces.  
Conversely, non-orientedness of the surfaces is crucial for generating a non-vanishing topological winding number. 
Therefore, the global non-orientedness of the vortex surfaces is necessary to generate a non-vanishing Pontryagin index originating from the vortex configuration.
Moreover, it is well known in topology that the number of intersection points of two closed two-dimensional surfaces in $\mathbb{R}^4$ is even. 
This implies that the number of self-intersection points of a closed surface is even, because the self-intersection number is defined by simply intersecting the surface with another surface infinitesimally displaced from it, i.e. a framing of the surface \cite{ER00}. 
Thus, even if each self-intersection point of an center vortex surface configuration gives a contribution $\pm 1/2$ to the Pontryagin index, the number of such contributions is even and the Pontryagin index is integer-valued. 
Moreover, they have pointed out that a non-zero Pontryagin index requires the existence of magnetic current. 
In the lattice study \cite{BFGO99}, in fact, the orientability of the vortex surfaces was studied, with result that these surfaces are non-orientable and have non-trivial genus in the confinement phase. 
These considerations connect  the gauge-invariant magnetic monopole and  the meron by way of a vortex and they become  topological objects simultaneously responsible for confinement.

\section{$N$-ality dependence of asymptotic string tension} 

Now we show that the fractional charge quantization is crucial to understand the $N$-ality dependence of asymptotic string tension.
The asymptotic string tension depends only on the $N$-ality of the quark charge (or the group representation of the Wilson loop). 
\footnote{
The $N$-ality $k$ of a given representation is obtained by the number of boxes in the corresponding Young diagram, mod $N$.
If $M(g)$ is the matrix representation of a group element $g$ in a representation of $N$-ality $k$, and $z \in \mathbb{Z}_N$ is an element in the center, then 
$M(z g)=\exp (2\pi i nk/N) M(g)$.
} 
This is because particles of zero $N$-ality can never bind to a particle of non-zero $N$-ality to form a color singlet and hence they can never break the string connecting two non-zero $N$-ality sources.  

Gluons belong to the adjoint representation of the gauge group and they have zero $N$-ality, so that all center elements are mapped to the identity. 
Therefore, gluons can not break the string formed between a pair of quark and antiquark in the $\bf{N}$ and $\bf{N}^*$ representations of $SU(N)$. 
Only other quarks, or other particles in the non-zero $N$-ality representations, can do that.  
The Wilson criterion for quark confinement should be understood to imply the  linear potential  rising indefinitely  in the limit in which the masses of any matter particles of non-zero $N$-ality are taken to infinity.

For example, in $SU(2)$ gauge theory, the center group is $\mathbb{Z}_2$ and the representations can be divided into two classes: $J=\frac12, \frac32, \cdots $ (half-integer) with $N$-ality one, and $J=1,2,\cdots$ (integer) with $N$-ality zero. The asymptotic string tension of $N$-ality=0 ($J=1,2,\cdots$) must be zero, while the non-vanishing asymptotic string tension of all $N$-ality=1 ($J=\frac12, \frac32, \cdots $) must be the same: 
\begin{equation}
 \sigma_J = \begin{cases}
  \sigma_{1/2} & (J=\frac12, \frac32, \cdots) \cr
  0 & (J=1, 2, \cdots) 
  \end{cases}
   .
\end{equation}
This expected result is easily derived from the considerations given in this paper:
The vortex contribution to the Wilson loop operator in the representation $J$ reads
\begin{align}
\exp \left[ igJ \int_{\Sigma} f(x;S) \right]
  = \exp [ iJ g\Phi L(C, S=\partial V) ]
  = z^{L(C, S=\partial V)}
,
\end{align}
where 
\begin{align}
 z := e^{iJg\Phi} 
= e^{2\pi J i n}
= \begin{cases}
  \pm 1 & (J=\frac12, \frac32, \cdots) \cr
  1 & (J=1, 2, \cdots) 
  \end{cases}
 , 
\end{align}
provided that the magnetic flux or the magnetic charge obeys the same fractional quantization condition (\ref{qc}). 
For $N$-ality=0, $z$ is trivial and the string tension is zero. For $N$-ality=1, $z \in \mathbb{Z}_2$ has the non-trivial value $-1$ when the vortex sheet $S$ intersects the world sheet $\Sigma$ and the string tension is non-zero. 
Thus the above argument reproduces correctly the $N$-ality dependence of the asymptotic string tension. 
The $SU(3)$ case will be discussed later.

\section{Consistency with lattice magnetic charge}\label{sec:lattice}

We consider how the above argument is consistent with the results of numerical simulations on a lattice, e.g., the magnetic monopole dominance in the string tension \cite{IKKMSS06}. 
For this purpose, it is instructive to recall the definition of  the magnetic charge and the magnetic current on a lattice for U(1) gauge theory \cite{Rothe05}. 
The basic idea is: if a magnetic monopole is located inside an elementary spatial cube on the lattice, then the enclosed magnetic charge can be determined by measuring the total magnetic flux through the surface (a set of plaquettes) of this cube.  For instance, the magnetic flux $\Phi_{jk}(x)$ through the surface $P_{x,jk}$ of a plaquette with an area $\epsilon^2$ lying in the $jk$-plane is related to the phase of the plaquette variable $U_{x,jk}=\exp [ig\epsilon^2 F_{jk}(x)]=\exp[ig\Phi_{jk}(x)]$
with $F_{jk}$ being the component of the magnetic field in the direction perpendicular to the $jk$-plane. 

The plaquette variable $U_{P}=\exp[ig\Phi_{P}]$ is given by the product of the four oriented link variables $U_{\ell}=\exp[i\theta_\ell] \in U(1)$ around the boundary of a plaquette $P$: $g\Phi_{P}=\sum_{\ell \in \partial P} \theta_\ell$ where $-\pi < \theta_\ell <\pi$. 
Hence, the phase (angle) associated with a given plaquette variable $U_{P}=\exp[ig\Phi_{P}]$ must satisfy
\begin{equation}
 -4\pi<g\Phi_{P} <4\pi 
 .
 \label{range}
\end{equation}
However, it should be remarked that the plaquette variable $U_{P}=\exp[ig\Phi_{P}] \in U(1)$ is a periodic function of $g\Phi_{P}$ with a period $2\pi$.
Therefore, the physical flux should be determined from $g\Phi_{P}$, modulo $2\pi n$, since the plaquette value remains unchanged by shifting $g\Phi_{P}$ by a multiple of $2\pi$. 
According to DeGrand and Toussaint \cite{DT80}, we decompose this angle into two parts:
\begin{equation}
 g\Phi_{P}=  2\pi n_{P}  + g\bar{\Phi}_{P} , 
 \quad n_{P} = 0, \pm 1, \pm 2, 
 \quad -\pi < g\bar{\Phi}_{P} < \pi  
 ,
\end{equation}
which covers the whole range of $\Phi_{P}$ (\ref{range}). 
If we add up the plaquette angles $\Phi_{P}$ of the six plaquettes bounding an elementary cube $c$, we will obtain a vanishing result, since each link is common to two plaquettes which give rise to the sum of two phases of the common link variable, equal in magnitude but of opposite sign, i.e.,
$\sum_{P \in \partial c} \Phi_{P}=0$. 
Therefore, the magnetic charge $Q_m$ as the magnetic flux through the closed surface $S=\partial c$ bounding the elementary cube $c$ is given by
\begin{equation}
  Q_m =  \sum_{P \in S=\partial c} \bar{\Phi}_{P}
  = - \sum_{P \in S=\partial c} \frac{2\pi}{g} n_{P} 
 .
\end{equation}
Thus the magnetic charge is a multiple of $\frac{2\pi}{g}$.
If a magnetic monopole is located in an elementary cube, then at least one of the plaquette angle must be larger in magnitude than $\pi$, so that there is a Dirac string (line) crossing the corresponding surface. 
The Dirac string can be moved around by making a large gauge transformation such that a particular link variable is mapped out of the principal value $[-\pi,\pi]$. 
But the net number of such strings leaving the elementary volume will not be affected by the gauge transformation.  Moreover, the number of magnetic monopoles contained in a volume $V$ is given by the sum of the magnetic monopole numbers of the elementary cubes making up the volume $V$. 

Next, we consider a lattice expression for the components of the magnetic current $k_\mu$ defined by 
$k_\mu=\frac12 \epsilon_{\mu\nu\rho\sigma}\partial_\nu F_{\rho\sigma}$. 
For example, the first component of the current $\bm{k}:=(k_1,k_2,k_3)$ is written in the form: 
\begin{equation}
   k_1 = \partial_2 F_{34} + \partial_3 F_{42} + \partial_4 F_{23} = (\partial_2,\partial_3,\partial_4) \cdot (F_{34},F_{42},F_{23})
 .
\end{equation}
Therefore, integrating $k_1$ over the volume of an elementary cube with edges along the 2,3 and 4 directions is equivalent to computing the flux of $\bm{F}:=(F_{34},F_{42},F_{23})$ through the surface of this cube due to the Gauss theorem:
$\int_{c}dx_2 dx_3 dx_4  k_1=\int_{S=\partial c} d\bm{S} \cdot \bm{F} =\sum_{P \in S=\partial c} \bar{\Phi}_{P} = - \sum_{P \in S=\partial c} \frac{2\pi}{g} n_{P}$.  
In order to compute the three components of $\bm{k}$, we need to calculate the flux through the plaquettes $P$ of three cubes having one link directed along the 4-axis. 
This computation is carried out in a completely analogous way as described in the above for the magnetic flux.  
Hence, by construction, each component of the magnetic current $k_\mu^L(x)$ on the lattice will be multiples of $\frac{2\pi}{g}$ when measured in lattice units: 
\begin{equation}
   k_\mu^L(x) 
= \frac{2\pi}{g} \frac12 \epsilon_{\mu\nu\rho\sigma} \partial_\nu n_{\rho\sigma}(x)
= \frac{2\pi}{g}  \partial_\nu {}^*n_{\mu\nu}(x)
 , 
 \quad n_{\rho\sigma}(x) = 0, \pm 1, \pm 2 
 ,
\end{equation}
where $\partial_\nu$ is the (forward) lattice derivative 
$\partial_\mu f(x):=[f(x+\epsilon \hat{\mu})-f(x)]/\epsilon$.  The magnetic current satisfies the topological conservation law $\partial_\mu^\prime k^L_\mu(x)=0$ and constitutes the closed loop in the dual lattice where $\partial_\mu^\prime$ is the backward lattice derivative defined by $\partial_\mu^\prime f(x):=[f(x)-f(x-\epsilon \hat{\mu}))]/\epsilon$.

Thus, it happens that the definition of the magnetic monopole due to DeGrand and Toussaint for the U(1) gauge theory is consistent with the observation made for the vortex in the continuum formulation for SU(2). 
The gauge-invariant magnetic charge defined on a lattice \cite{IKKMSS06} reduces to the above one due DeGrand and Toussaint  by taking a special gauge. Hence the result of \cite{IKKMSS06} is consistent with the observation made in this paper.

\section{$SU(3)$ case} 

 In the case of $SU(N)$ gauge group, it has been shown \cite{Kondo08} that the magnetic charge $q_m$ measured by the Wilson loop is subject to the quantization condition:  
\begin{equation}
  q_m =  \frac{2\pi}{g} \sqrt{\frac{2N}{N-1}} n , \  n \in \mathbb{Z} 
  ,
\end{equation}
which is analogous to the Dirac type, but different from it. 
Suppose that the magnetic flux $\Phi$ obeys the fractional quantization condition:
\begin{equation}
  \Phi = f_N \frac{2\pi}{g} \sqrt{\frac{2N}{N-1}} n , \  n \in \mathbb{Z} 
  .
\end{equation}
Then its  contribution to the Wilson loop is written as 
\begin{align}
\exp \left[ ig \sqrt{\frac{N-1}{2N}} \int_{\Sigma} f(x;S) \right]
 = \exp \left[ i \sqrt{\frac{N-1}{2N}} g \Phi L(C, S=\partial V) \right]
  = z_{N}^{L(C, S=\partial V)}
, 
\end{align}
where
\begin{align}
 z_{N} :=    \exp \left\{ i \sqrt{\frac{N-1}{2N}} g \Phi  \right\} 
= e^{2\pi i f_N n} 
 .
\end{align}
The center of $SU(N)$ is $\mathbb{Z}_N$. 
For $z_{N} {\bf 1}$ to belong to the center of SU(3), i.e., $\mathbb{Z}_3 \ni \{  {\bf 1}, e^{2\pi i/3} {\bf 1}, e^{4\pi i/3} {\bf 1}=e^{-2\pi i/3} {\bf 1} \}$, $f_3$ must take the fractional number except for a trivial case $f_N=0$:
\begin{equation}
  f_3 =  \frac13, \quad  \frac23 
  ,
\end{equation}
just as  the aforementioned $SU(2)$ case with the center subgroup  $\mathbb{Z}_2 \ni \{  {\bf 1},   e^{\pi i} {\bf 1}=- {\bf 1} \}$,
\begin{equation}
  f_2 =  \frac12   
  .
\end{equation}
This may be related to the fact that each self-intersection point of a vortex surface contributes $\pm k/N$ ($k= 1, \cdots, N-1$) to the Pontryagin index \cite{ER00}.

Finally, we consider the relationship between center vortices and our gauge-invariant vortices for $SU(N)$ case. 
We introduce $\mathcal{H}$ given by \cite{Kondo08}
\begin{equation}
 \mathcal{H} := 
\bm{\Lambda} \cdot \bm{H}  = \sum_{j=1}^{r} \Lambda_j H_j 
 ,
\end{equation}
where $H_j$ ($j=1,, \cdots, r$) are the generators from the Cartan subalgebra ($r=N-1$ is the rank of the gauge group $G=SU(N)$) and 
$r$-dimensional vector $\Lambda_j$ ($j=1,, \cdots, r$) is the highest weight of the representation in which the Wilson loop is considered. 

For SU(2), every representation is specified by a  half integer $J$:
\begin{equation}
 \Lambda_1 =J = \frac12 , 1, \frac32, 2, \cdots 
  , \quad
  H_1=\frac{\sigma_3}{2}
   ,
\end{equation}  
and the fundamental representation $J=\frac12$ of $SU(2)$ leads to 
\begin{equation}
 \mathcal{H}  
 = \frac{1}{2} {\rm diag} \left( \frac{1}{2}, \frac{-1}{2} \right) 
= \frac12 \frac{\sigma_3}{2}  
  .
\end{equation}

For SU(3), the highest weight of the representation characterized by the Dynkin indices $[m,n]$ is given by 
\begin{equation}
 \bm{\Lambda} 
 = (\Lambda_1,\Lambda_2) 
=  \left({m \over 2}, {m+2n \over 2\sqrt{3}} \right) 
 .
\end{equation}
The explicit form of $\mathcal{H}$  for the fundamental representations $[m,n]$ reads using the diagonal set of the Gell-Mann matrices $\lambda_3$ and $\lambda_8$:
\begin{align}
 \mathcal{H} 
=  \frac{1}{2} (\Lambda_1 \lambda_3 + \Lambda_2 \lambda_8)
= \frac{1}{2} {\rm diag} \left( \frac{2m+n}{3}, \frac{-m+n}{3} , \frac{-m-2n}{3} \right)
 ,
\end{align}
and we enumerate all fundamental representations $\textbf{3}$:
\begin{subequations}
\begin{align}
[m,n]=[1,0]: 
\bm{\Lambda} 
=& \left( \frac{1}{2}, \frac{1}{2\sqrt{3}} \right) := \nu_1 
 , \quad
  \mathcal{H} = \frac{1}{2} {\rm diag} \left( \frac{2}{3}, \frac{-1}{3} , \frac{-1}{3} \right) 
   ,
   \label{fr1}
\\
[m,n]=[-1,1]: 
 \bm{\Lambda} 
=& \left( \frac{-1}{2}, \frac{1}{2\sqrt{3}} \right) := \nu_2 
 , \quad
  \mathcal{H} = \frac{1}{2} {\rm diag} \left( \frac{-1}{3}, \frac{2}{3} , \frac{-1}{3} \right)
   ,
   \label{fr2}
\\
[m,n]=[0,-1]: 
 \bm{\Lambda} 
=& \left( 0, \frac{-1}{\sqrt{3}} \right) := \nu_3 
 , \quad
  \mathcal{H} = \frac{1}{2} {\rm diag} \left( \frac{-1}{3}, \frac{-1}{3} , \frac{2}{3} \right)
  = \frac{-1}{\sqrt{3}} \frac{\lambda_8}{2} 
   ,
   \label{fr3}
\end{align}
\end{subequations}
and their conjugates $\textbf{3$^*$}$.
Therefore, the center vortex is obtained by replacing the magnetic flux $\Phi$ with the diagonal matrix $4\pi \mathcal{H}$:
\begin{equation}
   \sqrt{\frac{N-1}{2N}}  g\Phi \rightarrow  4\pi \mathcal{H} 
  ,
\end{equation}
which indeed leads to the non-trivial elements of the center group $\mathbb{Z}_N$:
\begin{align}
z_{N} =&    \exp \left\{ i \sqrt{\frac{N-1}{2N}} g \Phi  \right\} 
\nonumber\\
  \rightarrow & \exp (4\pi i \mathcal{H}) =  e^{2\pi i f_N} {\bf 1}
\in \mathbb{Z}_N , \quad
   f_N = k/N \ (k= 1, \cdots, N-1)
 ,
\end{align}
where $f_N=0$ corresponds to the trivial element ${\bf 1}$.
Thus the center vortices are replaced by our gauge-invariant vortices carrying the fractional magnetic flux $f_N \Phi$. 

If this observation is correct, the magnetic part of the $SU(N)$ Wilson loop with an additional fractional factor $f_N$:
\begin{equation}
 W_C^m    
= \exp \left\{ i g f_N \sqrt{\frac{N-1}{2N}} (k, \Xi_{\Sigma})  \right\}
 , \quad
    f_N = k/N \ (k= 1, \cdots, N-1) 
\end{equation}
will reproduce the string tension for any $k$ ($k= 1, \cdots, N-1$), just as confirmed for $SU(2)$ gauge group in \cite{IKKMSS06}.  
This is a conjecture derived in this paper. 
This is consistent with the center vortex mechanism for quark confinement. However, it will be rather difficult to identify the vortex structure in $SU(3)$ case in numerical simulations. For the vortex surfaces for SU(N), $N \ge 3$, may branch and the superimposed magnetic fluxes in general also modify the type of vortex flux, i.e., its direction in color space.

\section{Conclusion and discussion} 

In this paper, we have given a gauge-invariant definition of the vortex surface $S$ such that the gauge-invariant magnetic monopole emerge in the boundary of the surface $S$. 
We do not use any gauge fixing for obtaining the relevant vortex and magnetic monopole, such as maximal Abelian gauge and maximal center gauge which are extensively used in the conventional approaches.   These results are obtained from the non-Abelian Stokes theorem for the Wilson loop operator \cite{Kondo08}.

For the topological defects such as magnetic monopoles or vortices to give a non-trivial contribution to the Wilson loop average, thereby yielding the area law, we have shown that the magnetic current must carry the fractional magnetic flux (charge) compared with that dictated by the ordinary quantization condition. 
Under this identification, our gauge-invariant vortices can play the same role as the center vortices in giving a non-trivial contribution to the Wilson loop average.  Hence, the gauge-invariant vortices are expected to reproduce a number of nice results obtained so far by center vortices \cite{Greensite03}. 

For center vortices,  it was pointed out \cite{ER00} that the global non-orientedness of the center vortex surface is necessary to generate a non-vanishing Pontryagin index from the vortex configuration in $D=4$ dimensional spacetime. 
The existence of the fractional magnetic flux is consistent with the non-orientedness of the closed vortex surface in $D=4$  where  the smooth closed vortex surface must consist of the surface patches of different orientations so that the magnetic flux in the boundary of an isolated patch inevitably becomes fractional.

In this paper, however, we have considered the infinitely thin magnetic flux tube.  Therefore, the asymptotic string tension is correctly understood to reproduce the correct $N$-ality dependence of the string tension.  In order to see the behavior of the string tension in an intermediate range of distance, e.g., Casimir scaling or Sine-law scaling of the $k$-string tension,  we need to consider the magnetic flux tube with a finite thickness at around 1 fermi and the associated thick vortex.  
This issue will be discussed in a subsequent paper.

Finally, we consider what happens to vortices as the number of colors $N$ is increased.  
It is known \cite{Witten79} that instanton effects (of order $\exp(-1/g^2)$) vanish exponentially $\exp(-N)$ for large $N$, since $g^2$ is of order $1/N$, or $\lambda:=g^2N$ is to be fixed in the large $N$ expansion. 
Therefore, instanton gas disappears in the large $N$ limit and predictions that depend on thinking about instantons and an instanton gas (even a dense gas) will not be correct.  
The argument for quantization of the topological charge $Q_P$ of instanton in QCD starts with  the boundary condition: 
the gauge field approaches a pure gauge at infinity,  
$
 \mathscr{A}_\mu \rightarrow G \partial_\mu G^{-1}
$
as $|x| \rightarrow \infty$. 
Thus, in a quark confining theory this boundary condition is not reasonable and the conclusion based on it are likely to be wrong, provided that QCD remains a confining theory as $N \rightarrow \infty$, as argued by 't Hooft \cite{tHooft74}.   
Do vortices get suppressed like instantons?  If so, how does one explain confinement using vortices for large $N$. 
We can argue that vortices are not suppressed even in the large $N$. 
According to \cite{Witten79}, quark confinement tells us that the vacuum cannot be regarded as being mostly pure gauge.  In fact, the expectation value of the Wilson loop in any state that is mostly pure gauge will (just as in perturbation theory) not show a confining potential.  
For any approximation that yields confinement will have to include fields that fluctuate at infinity, corresponding to a vacuum that is not mainly pure gauge.  
As argued in the text, our vortex picture is accompanied by  merons with half-integral topological number $Q_P$ and the meron field is written as $
 \mathscr{A}_\mu = \frac12 G \partial_\mu G^{-1}
$, which is not the pure gauge. 
Therefore, the existence of vortices are not in conflict with the large $N$ expansion. 
This suggests that vortex gas will not be suppressed and survive in the large $N$ limit to be responsible for confinement.  More details will be given elsewhere.

\section*{Acknowledgments}
The author would like to thank Seikou Kato for reading the manuscript and for helpful discussions. 
He is appreciative of continuous  discussions from Akihiro Shibata, Takeharu Murakami and Toru Shinohara.
This work is financially supported by Grant-in-Aid for Scientific Research (C) 18540251 from Japan Society for the Promotion of Science
(JSPS).


\end{document}